\documentclass[10pt]{article}
\usepackage{graphicx}
\usepackage{amssymb}
\usepackage{amsmath}
\usepackage{caption}
\usepackage{subcaption}
\usepackage{listings}
\usepackage{url}
\usepackage{enumitem}
\usepackage{color}   
\usepackage{hyperref}
\usepackage[margin = 0.9in]{geometry}
\usepackage[section]{placeins}
\usepackage{authblk}
\hypersetup{
    colorlinks=false, 
    linktoc=all,     
    linkcolor=blue,  
}

\newif\ifarXiv
\newif\ifSnP
\arXivtrue

\graphicspath{ {./images/} }

\begin{document}

\title{An operational architecture for privacy-by-design in public service applications}
\author[1]{\href{http://www.cse.iitd.ac.in/\~prashant}{Prashant Agrawal}}
\author[2]{\href{https://www.dvara.com/research/our-team/anubhutie-singh/}{Anubhutie Singh}}
\author[2]{\href{https://www.dvara.com/research/our-team/malavika-raghavan/}{Malavika Raghavan}}
\author[1,3]{\href{http://www.cse.iitd.ac.in/\~svs}{Subodh  Sharma}}
\author[1,3]{\href{http://www.cse.iitd.ac.in/\~suban}{Subhashis Banerjee}}
\affil[1]{\href{http://www.cse.iitd.ac.in}{Department of Computer Science and Engineering}, \href{http://www.iitd.ac.in}{IIT Delhi}}
\affil[2]{\href{https://www.dvara.com/research/}{Dvara Research}}
\affil[3]{Also associated with the School of Public Policy, IIT Delhi}

\renewcommand\Authands{ and }

\date{\today}
\pagecolor{white}  

\maketitle

\begin{abstract}
Governments around the world are trying to build large  data registries for effective delivery of  a variety of public services. However, these efforts are often undermined due to  serious concerns over privacy risks associated with collection and processing of personally identifiable information. While a rich set of special-purpose privacy-preserving techniques exist in computer science, they are unable  to provide end-to-end protection in alignment with  legal principles in the absence of an overarching operational architecture to ensure purpose limitation and protection against insider attacks. This either leads to weak privacy protection in large designs, or adoption of overly defensive strategies to protect privacy by compromising on utility. 

In this paper, we present an operational architecture for privacy-by-design based on independent regulatory oversight stipulated by most data protection regimes, regulated access control, purpose limitation and data minimisation. We briefly discuss the feasibility of implementing our architecture based on existing techniques. We also present some sample case studies of privacy-preserving design sketches of challenging public service applications.
\end{abstract}

\section{Introduction}
\label{sec:intro}

A welfare state may have legitimate interests in building large data registries with personally identifiable information (PII) for efficiency of service delivery. A state may also legitimately need to put  its residents under purpose-specific surveillance. In fact, several commentators have alluded to the possibility of pervasive under-the-skin
surveillance in a post-COVID world \cite{Harari}. 
However, mandatory recordings of PII require enacting  reasonable and fair laws to ensure that the processing of PII is proportionate to the stated objective, 
and safeguard the basic operative principles of privacy and fairness. Citizens' basic rights need to be protected  even when there is a legitimate state interest in digitisation with PII \cite{puttaswamy}.  The need to ensure that the information collected is not used adversely against citizens to harm them takes us into one of the hard problems of modern public policy: creating rules and technologies around information privacy to help strike this critical balance for online collection of PII at national scale. 

In this paper we address the problem of operationalising  the broad privacy-by-design principles outlined in \cite{pbd,oecd}, in the context of large public service databases. We present an architecture for implementing the data protection principles after the utility and proportionality of an application  have been established through an appropriate regulatory analysis \cite{solove,gdpr,pdpbill}.  

The general principles of fair and reasonable processing, purpose, collection and storage limitation, notice and consent, data quality etc. have evolved since the 1970s, both through sector specific standards in the US such as the Social Security Number Protection Act \cite{ssn} and Health Insurance Portability and Accountability Act (HIPAA)  \cite{hipaa}, or through omnibus laws in general data protection standards such as the GDPR in the European Union \cite{gdpr} and the Draft Data Protection Bill of India \cite{pdpbill}. However, they have largely failed to prevent both direct harms that can occur  as a result of data breaches or  through unauthorised access of personal data - such as identity thefts, unethical profiling and unlawful surveillance, or secondary harms that could  arise  due to the use of the data to adversely affect a person - such as through discrimination or exclusion, predatory targeting for unsuitable products, loss of employment, inaccurate credit rating etc.  Dictums such as personal data shall be processed in a fair and reasonable manner are non-specific, and they do not adequately define the contours of the required regulatory  actions. As episodes like Cambridge Analytica \cite{cambridge-analytica} demonstrate, harm is often not immediately obvious, and causal links of harm are not always easy to determine. This is compounded by the fact that data collection and use are becoming ubiquitous making it hard to trace misuse; the effects of misuse of personal data may not immediately manifest, and when they do they may not be easily quantifiable in monetary terms despite causing grave distress. Hence, ex-post accountability and punitive measures are largely ineffective, and it is imperative to operationalise  ex-ante preventive principles. 

As a consequence of the weak  protection standards, most attempts at building large public services like national identity systems \cite{lsereport,SSS},  health registries \cite{caredata,australia,healthrec}, national population and voter registries \cite{voter1,voter2,voter3}, public credit registries \cite{equifax,dvara},  income  \cite{income} and tax registries \cite{irsprivacy} etc. have often been questioned on privacy and fairness grounds and have been difficult to operationalise.  The concerns  have invariably been related to the need for protective safeguards when large national data integration projects are contemplated by  governments and acknowledgment of the unprecedented surveillance power that this could create. In some situations they have even had to be abandoned altogether  as they were unable to deal with these risks \cite{caredata,australia,ukidentity}. In India too, the recent momentum and concerns around informational privacy guarantees have occurred in the context of the creation of new government databases and digital infrastructures for welfare delivery \cite{aadhaar,DoA}.  

\subsection{Requirements for privacy protection}

Recording transactions with PII projects an individual into a data space, and any subsequent loss of privacy can happen only through the data pathway. Hence data protection is central to privacy protection insofar as databases are concerned. The critical challenge in design of a data protection framework is that the main uses of digitisation - long term record keeping and data analysis - are seemingly contradictory to the privacy protection requirements. 
The legal principles around ``fair information practice'' attempt to reconcile these tensions,
but there are four broad areas that require careful attention for effective data protection.

\subsubsection{Use cases and data minimisation}
First, a data protection framework is incomplete without an investigation of the nuances of digital identity, and guidelines for the various use cases of authentication, authorisation and accounting. It is also incomplete without an analysis of the extent to which personal information needs to be revealed for each use case, for example during know-your-citizen or -customer (KYC) processes. In addition, effective protection requires an understanding of the possible pathways of information leaks; of the limits of anonymisation with provable guarantees against re-identification attacks \cite{robust-deanon}; and of the various possibilities with virtual identities \cite{security-without-identification,vid}.

\subsubsection{Access control and purpose limitation}
Second, there have to be clear-cut guidelines for defining the requirements and standards of access control, and protection against both external and insider attacks in large data establishments,  technically as well as legally. In particular, insider attacks are the biggest threat to privacy in public databases \cite{SSS}. These include possible unauthorised and surreptitious examination of data, transaction records, logs and audit trails by personnel with access, leading to profiling and surveillance of targeted groups and individuals, perhaps at the behest of interested and influential parties in the state machinery itself \cite{insider}. Thus, there must be guidelines on how the data may be accessed, under what authorisation and for what purpose. In addition, post data access purpose limitation - ensuring that there is no illegal use after the data crosses the access boundaries -  is also crucial for privacy protection.

\subsubsection{Inferential privacy and purpose limitation}
Third, a data protection framework is incomplete without guidelines for safe use of AI and data analytics. Most theories for improving state efficiency in delivery of welfare and health services using personal data will have to consider improved data processing methods for targeting, epidemiology, econometrics, tax compliance, corruption control, analytics, topic discovery, etc. This, in turn, will require digitisation, surveillance and processing of large-scale personal transactional data.  This requires detailed analyses of how purpose limitation of such surveillance - targeted towards improving efficiency of the state's service delivery - may be achieved without enabling undesirable mass surveillance that may threaten civil liberty and democracy. There must also be effective guidelines to prevent discriminatory and biased data processing \cite{fairml-book}.

\subsubsection{Regulatory oversight}
Finally, it is well recognised in data protection frameworks \cite{solove,gdpr,pdpbill} that regulatory oversight is a necessary requirement for ensuring the above. 

\subsection{Our contribution}
While there is a rich set of tools and techniques in computer science arising out of decades of innovative privacy research, there is  no overarching general framework for a privacy preserving architecture which, in particular, allows regulatory supervision and helps deal with the above issues in effective designs.
In this paper we propose such an operational architecture for implementing the data protection principles. Our immediate objective here is design space exploration and not specific implementations to evaluate performance and scalability.

We illustrate the effectiveness of our proposal through design sketches of some challenging large public service applications. In particular, we illustrate through some real world case studies how some state-of-the-art designs either fail in their data protection goals, or tend to be overly defensive at the cost of utility in the absence of such an architecture. 

The rest of the paper is organized as follows. Section \ref{sec:context} briefly reviews
\ifarXiv
 the basic legal principles  for data protection. 
 \fi
 Section \ref{sec:privacy-in-cs} reviews concepts, tools and techniques from computer science for privacy protection. 
Section \ref{sec:trusted-executables} presents our operational architecture.
Section \ref{sec:feasibility} discusses the feasibility and Section \ref{sec:case-studies} discusses some illustrative case studies of large government applications.

\ifarXiv

\section{The India context and privacy and data protection concepts in law and regulation}
In what follows we briefly discuss the context of digitisation and privacy in India and the basic legal principles around privacy. We situate this analysis within the context of India's evolving regulatory and technical systems. However, many of these principles are relevant for any country seeking to align legal and technical guarantees of privacy for citizens.

\label{sec:context}
\subsection{Public digital infrastructures and informational privacy in India}
\label{sec:India}
Building public digital infrastructures has received an impetus in India in recent times \cite{cacm-indiadigital} and privacy has been an obvious concern.
India has a long-standing legal discourse on privacy as a right rooted in the country's Constitution. However, informational privacy and data protection issues have gained renewed visibility due to the recent national debate around the country's Aadhaar system \cite{aadhaar}. Aadhaar is a unique, biometric-based identity system launched in 2009, with the ambitious aim of enrolling all Indian residents, and recording their personal information, biometric fingerprints and iris scans against a unique identity number. 
Aadhaar was designed as a solution for preventing leakages in government welfare delivery and targeting public services through this identity system.  In addition, the ``India stack'' was envisioned as a set of APIs that could be used - by public and private sector entities contract - to query the Aadhaar database to provide a variety of services \cite{cacm-indiastack}. However, as the project was unrolled across the country, its constitutionality was challenged in the courts on many grounds including the main substantive charge that it was violative of the citizens' right to privacy. Over 30 petitions challenging the system were eventually raised to the Supreme Court of India for its final determination. In the course of the matter, a more foundational question arose, i.e., whether the Indian Constitution contemplated a fundamental right of privacy? 
The question was referred to a separate 9-judge bench of the Indian Supreme Court to conclusively determine the answer to this question. The answer to this question is important both for law and computer science, since the response creates deep implications for the design of technical systems in India. The Supreme Court's unanimous response to this question in Justice K.S.Puttaswamy (Retd.) vs Union of India  (Puttaswamy I) \cite{puttaswamy} was to hold that privacy is a fundamental right in India guaranteed by Part III (Fundamental Rights) of the Indian Constitution. 
Informational privacy was noted to be an important aspect of privacy for each individual, that required protection and security. In doing so, the Court recognised the interest of an individual in controlling or limiting the access to their personal information, especially as ubiquitous data generation and collection, combined with data processing techniques, can derive information about individuals that we may not intend to disclose.

\subsection{Defining informational privacy} 
\label{sec:law}
In addition to cementing privacy as a constitutional right for Indians, the Supreme Court in Puttaswamy I \cite{puttaswamy} also played an important role in clarifying certain definitional aspects of the concept. 

First, when defining privacy, the lead judgement noted that every person's reasonable expectation of privacy has both subjective and objective elements (see page 246 of Puttaswamy I), i.e., 
\begin{enumerate}
\item the subjective element which is to the expectation and desire of an individual to be left alone, and
\item	the objective element, which refers to objective criteria and rules (flowing from constitutional values) that create the widely agreed content of ``the protected zone'', where a person ought to be left alone in our society. 
\end{enumerate}

Second, informational privacy was also recognised (see page 201 of Puttaswamy I, from a seminal work which set out a typology of privacy) to be:

\begin{quote}
``$\ldots$ an interest in preventing information about the self from being disseminated and controlling the extent of access to information.''
\end{quote}

It would be the role of a future Indian data protection law to create some objective standards for informational privacy to give all actors in society an understanding of the ``ground rules'' for accessing an individuals' personal information. These principles are already fairly well-developed through several decades of international experience. India is one of the few remaining countries in the world that is yet to adopt a comprehensive data protection framework.  This section provides a brief overview of some of these established concepts. 

\subsection{Data protection principles}

One of the early and most influential global frameworks on privacy protection are the 1980 OECD Guidelines on the Protection of Privacy and Transborder Flows of Personal Data \cite{oecd}. These were formulated as a response to the advancements in technology that enabled faster processing of large amounts of data as well as their transmission across different countries.   These Guidelines were updated in 2013, reflecting the multilateral consensus of the changes in the use and processing of personal data in that 30 year period. Therefore, it is a good starting point for the fundamental principles of privacy and data protection.

The key principles of the OECD Privacy Framework 2013  are:

\begin{description}
\item[Collection Limitation:] Personal data should be collected in a fair and lawful manner and there should be limits to its collection.
\item[Use Limitation:] Collected personal data be used or disclosed for any purposes other than those stated. If personal data must be used for purposes other that those stated, it should with the consent of the data subject or with the authority of the law.
\item[Purpose Specification:] The purpose for collection of personal data should be stated no later than the point of collection. All subsequent uses of such data must be limited to the stated purposes. 
\item[Data Quality:] Collected personal data should be relevant for the stated purposes and its accuracy for such a purpose must be maintained.
\item[Security Safeguards:] Reasonable safeguards must be adopted by the data controller to protect it from risks such as unauthorised access, destruction, use, modification or disclosure of the data.
\item[Accountability:] Any entity processing personal data must be responsible and held accountable for giving effect to the principles of data protection and privacy.
\item[Openness:] Any entity processing personal data must be transparent about the developments and practices with respect to the personal data collected.
\item[Individual Participation:] Individuals should have the rights to confirm from the data controller whether they have any personal data relating to them and be able to obtain the same within a reasonable time, at a reasonable charge and in a reasonable manner. If these requests are denied, individuals must be given the reasons for such denial and have the right to challenge such denials. Individuals must also retain the right to be able to challenge personal data relating to them and able to erase, rectify, complete or amended.
\end{description}

These principles, and many international instruments and national laws that draw from them, set some of the basic ground rules around the need for clear and legitimate purposes to be identified prior to accessing personal information. They also stress on the need for accountable data practices including strict access controls.  Many of these principles are reflected to varying degrees in India's Personal Data Protection Bill in 2019 \cite{pdpbill} which was introduced in the Lower House of the Indian Parliament in December 2019. The Bill is currently under consideration by a Joint Select Committee of Parliamentarians  following which it will enter Parliament for final passage. 

The OECD Privacy Framework 2013 \cite{oecd} in Article 19(g)  also recognised the need for the promotion of technical measures to protect privacy in practice. There is also a growing recognition that if technical systems are not built with an appreciation of data protection and privacy principles, they can create deficits of trust and other dysfunctions. These are particularly problematic in government-led infrastructures. 

\subsection{The failure of privacy self-management and the need for accountability-based data protection}

\fi

The need for data processing entities to adhere to objective and enforceable standards of data protection is heightened because of vulnerability of the individuals whose data they process. Although research shows that individuals value their privacy and seek to control how information about them is shared, cognitive limitations operate at the level of the individuals' decision-making about their personal data \cite{solove}. This ``Privacy Paradox'' signals the behavioural biases and information asymmetries that operate on people making decisions about sharing their personal information.  Especially in contexts where awareness that personal data is even being collected in digital interactions is low, such as with first-time users of digital services in India, it is often unfair and meaningless to delegate the self-management of privacy to users entirely through the ineffective mechanism of ``consent''. The inadequacy of consent alone as a privacy protection instrument  has been well established, especially given that failing to consent to data collection could result in a denial of the service being sought by the user \cite{solove}.  

In the context of these findings, it is crucial that digital ecosystems be designed in a manner that protects the privacy of individuals, does not erode their trust in the data collecting institution and does not make them vulnerable to different natures of harm. Therefore, mere dependence on compliance with legal frameworks by data controllers is not sufficient. Technical guarantees that the collected data will only be used for the stated purposes and in furtherance of data protection principles must become a reality, if these legal guarantees are to be meaningful. The need for early alignment of legal and technical design principles of data systems, such as access controls, purpose limitation and clear liability frameworks under    appropriate regulatory  jurisdictions are essential to create secure and trustworthy public data infrastructures \cite{solove, gdpr, pdpbill}. 

\section{Privacy concepts in computer science}
\label{sec:privacy-in-cs}

Before we present our architectural framework,  we briefly review some privacy preserving tools from computer science.

\subsection{Encryption, signatures and cryptographic hashes}

\subsubsection{Encryption}
\label{ssec:encryption}

Cryptographic encryption \cite{encryption}, for protecting data either in storage or transit, have often been
advocated for privacy protection. The following types are of particular importance:
\begin{description}
\item[Symmetric encryption] Symmetric encryption allows two parties
   to encrypt and decrypt messages using a shared secret key.  
  Diffie-Hellman key exchange protocol \cite{DH} is commonly  used by the parties to
  jointly establish a shared key over an insecure channel.
  
\item[Asymmetric encryption] Asymmetric or public key
  encryption \cite{DH} allows two parties to communicate
  without the need to exchange any keys beforehand. Each party holds a pair of {\em public} and {\em private}
  keys such that messages encrypted using the receiver's public key
  cannot be decrypted without the knowledge of the corresponding
  private key.

\item[ID-based encryption] ID-based encryption \cite{ibe} allows the
  sender to encrypt the message against a textual ID
  instead of a public key. A trusted third party provisions decryption
  keys corresponding to the IDs of potential receivers after
  authenticating them through an out-of-band mechanism. 
  ID-based encryption considerably simplifies the
  public key infrastructure: a sender can encrypt messages using the
  semantic identifier of the intended recipient without explicitly
  knowing the public keys of the particular receivers.
\end{description}

Encryption with strong keys is a powerful method for
privacy protection provided 
there are
no unauthorised accesses to the keys. Insider attacks, however, pose serious
 risks if the keys also
reside with the same 
authority. Even when the keys are
stored securely, they have to be brought into the memory for
decryption during run-time, and can be leaked by a compromised
privileged software, for example an operating system or a hypervisor.

\subsubsection{Signatures} 
\label{ssec:bsign}

\begin{description}
\item[Digital signature] A digital signature \cite{DH}
  $\sigma_{pk}(m)$ on a message $m$ 
  allows a verifier to verify using  the public key $pk$ that  $m$ was
  indeed signed with the corresponding the private key. Any alteration of $m$
  invalidates the signature. Signatures also provide
  non-repudiation. 
\item[Blind signatures] Blind signatures \cite{blind-sign} are
  useful 
  to obtain a signature on a
  message 
  without exposing the contents 
  of
  the message to the signer. 
  A signature $\sigma'_{pk}(b(m))$ by a signer holding
  public key $pk$ 
  allows the signer
  to sign a {\em blinded message} $b(m)$ that does not reveal anything about
  $m$. The  author of the message  can now use the  $\sigma'_{pk}(b(m))$ to 
  create an unblinded digital signature $\sigma_{pk}(m)$.
\end{description}

\subsubsection{Cryptographic hash function (CHF)}
CHFs are functions that are $a)$ `one-way', i.e., given hash value $h$, it is difficult to find an $x$ such that $h=hash(x)$, and $b)$ `collision-resistant', i.e., finding any $x_1$ and $x_2$ such that $hash(x_1)=hash(x_2)$ is difficult. CHFs form the basis of many privacy preserving cryptographic primitives.

\subsection{Data minimisation}
\label{sec:minimise}
There are several techniques from computer science that are
particularly useful for data minimisation - at different levels of
collection, authentication, KYC, storage and dissemination. Some of
these are:

\subsubsection{Zero knowledge proofs (ZKPs) and selective disclosures}
\label{ssec:zkp}

ZKPs \cite{zkp} are proofs that allow a party to
prove to another that a statement is true, \emph{without} leaking any
information other than the statement itself. Of particular relevance
are ZKPs of knowledge \cite{zkpk}, which convince a
verifier that the prover knows a secret without revealing it.
ZKPs also enable \emph{selective disclosure} \cite{selective-disclosure},
i.e., 
individuals can prove only purpose-specific
attributes about their identity without revealing additional
details; for example, that one is of legal drinking age without revealing the age itself. 

\subsubsection{Anonymity and anonymous credentials}

``Anonymity refers to the state of being not identifiable within a set
of individuals, the anonymity set'' \cite{anonterm}. In the context of
individuals making transactions with an organisation, the following
notions of anonymity can be defined:

\begin{description}
\item[Unlinkable anonymity] Transactions provide unlinkable anonymity
  (or simply \emph{unlinkability}) if $a)$ they do not reveal the true
  identities of the individuals to organisations, and $b)$ organisations
  cannot identify how different transactions map to the individuals. 

\item[Linkable anonymity] Transactions provide linkable anonymity if
  an organisation can identify whether or not two of its transactions involve
  the same individual, but individuals' true identities remain
  hidden. Linkable anonymity is useful because it allows individuals
  to maintain their privacy while allowing the organisation to
  aggregate multiple transactions from the same individual. Linkable
  anonymity is typically achieved by making individuals use
  \emph{pseudonyms}. 
\end{description}


\begin{description}
\item[Anonymous credentials]  Authenticating individuals online may require them to provide
credentials from a credential-granting organisation $A$ 
to a credential-verifying
organisation $B$. 
Privacy protection using  anonymous credentials \cite{anoncred,vid,security-without-identification} can ensure that transactions with $A$
are unlinkable to transactions with $B$. Anonymous credentials
  allow an
  individual to obtain a credential from an organisation $A$ against
  their pseudonym with $A$ and transform it to an identical credential
  against their pseudonym with organisation $B$. 
  An identity authority provisions a master identity to each
  individual from which all pseudonyms belonging to an individual,
  also known as \emph{virtual identities}, are cryptographically
  derived. Anonymous credentials are typically implemented by
  obtaining blind signatures (see Section \ref{ssec:bsign}) from the
  issuer and using ZKPs of knowledge (see Section
  \ref{ssec:zkp}) of these signatures to authenticate with the
  verifier. The credential mechanism guarantees:
\begin{itemize}
	\item \emph{Unlinkable anonymity across organisations.} 
          This property ensures that $A$ cannot
          track the uses of the issued credentials and $B$ cannot
          obtain the individual's information shared only with $A$ even
          when $A$ and $B$ collude.
	\item \emph{Unforgeability.} A credential against an
          individual's pseudonym cannot be generated without obtaining
          an identical credential against another pseudonym
          belonging to
          the same individual.
	\item \emph{Linkable anonymity within an organisation.}
          Depending on the use case requirements, individuals may  or may not use more than one pseudonym per
          organisation. In the latter case the transactions within an organisation also become unlinkable.
\end{itemize}
If an organisation $A$ requires to link multiple transactions from the
same individual, it can indicate this requirement to the identity
authority that checks if pseudonyms used by individuals with $A$ are
unique. If $A$ does not require linking, the identity authority merely
checks if the pseudonyms are correctly derived from the individual's
master identity. 
If the checks pass, an anonymous credential certifying this
fact is issued by the identity authority. All checks by the identity authority preserve individuals' anonymity.

\item[Accountable anonymous credentials] Anonymity comes with a price
  in terms of accountability: individuals can misuse their credentials
  if they can never be identified and held responsible for their
  actions. 
  Trusted third parties can revoke
  the anonymity of misbehaving users to initiate punitive measures
  against them
  \cite{anoncred-optional-revocation,anoncred-dynamic-revocation,anoncred-group-signature}. 
  {\em One-time credentials} and {\em $k$-times anonymous
  authentication schemes}
  \cite{ktime-anoncred,ktime-anoncred2,ktime-anoncred3} also prevent
  overspending of limited-use credentials by revoking individuals' anonymity if
  they overspend. 
  Blacklisting
  misbehaving users for future access without revoking their anonymity is also feasible
  \cite{blacklistable-anoncred}.

\item[Linkability by a trusted authority] 
  Linking across organisations may also be required for legitimate purposes, for example for legitimate data mining. 
  Also see examples in Section \ref{sec:feasibility}. Such  linkability  also seems to be an inevitable requirement to
  deter sharing of anonymous credentials among individuals
  \cite{linkability-for-credential-sharing}.
Linkability by a trusted authority can be trivially achieved by
individuals attaching a randomised encryption of a unique identifier
against the trusted authority's public key for transactions
requiring cross-linking. Of course, appropriate mechanisms must exist to ensure
that the trusted authority does not violate the legitimate purpose of
linking.
\end{description}

Note that the anonymity of credentials is preserved only under the
assumption that individuals interact with organisations through
anonymous channels (e.g., in
\cite{anoncred-optional-revocation}). In particular, neither the
communication network nor the data that individuals share with
organisations should be usable to link their transactions (see Section
\ref{sec:anonNw} and \ref{sec:anonymisation}). 

\subsubsection{Anonymous networks}
\label{sec:anonNw}
Anonymous networks, originally conceptualised as \emph{mix networks}
by Chaum \cite{mixnets}, are routing protocols that make messages
hard-to-trace. Mix networks consist of a series of
proxy servers where each of them receives messages from multiple
senders, shuffles them, and sends to the next proxy server. An
onion-like encryption scheme allows each proxy server to only see an
encrypted copy of the message (and the next hop in plaintext), thus
providing untraceability to the sender even if only one proxy server
honestly shuffles its incoming messages.

\subsubsection{Database anonymisation}
\label{sec:anonymisation}
Anonymisation is the process of transforming a database 
such that individuals' data cannot be traced back to
them. 
However, 
research in 
de-anonymisation has shown that anonymisation 
does not work in
practice, as small number of data points about individuals coming from
various sources, none uniquely identifying, can completely identify
them when combined together \cite{robust-deanon}. This is backed by
theoretical results \cite{anonymisation-high-dimension,
  provable-deanonymisation} which show that for high-dimensional data,
anonymisation is not possible unless the amount of noise introduced is
so large that it renders the database useless. There are several
reports in literature of de-anonymisation attacks on
anonymised social-network data
\cite{deanon-social-network-1,deanon-social-network-2}, location data
\cite{deanon-location}, writing style \cite{deanon-writing}, web browsing
data \cite{deanon-browsing-history}, etc.

\subsubsection{Interactive database anonymisation}
\label{sec:query-restrict}
In this setting, analysts interact with a remote server only through a
restricted set of queries and the server responds with possibly noisy
answers to them. Dinur and Nissim \cite{revealing-while-privacy}
show that given a database with $n$ rows, an adversary having no
prior knowledge could make ${O}(n \; \mbox{{\em polylog}}(n))$ random
subset-sum queries to \emph{reconstruct} almost the entire database,
unless the server perturbs its answers too much (by at least
$O(\sqrt{n})$). This means that preventing inference attacks 
is impossible if the adversary is allowed to make
arbitrary (small) number of 
queries. Determining whether a
given set of queries preserves privacy against such attacks is in
general intractable (NP-hard) \cite{auditing-boolean-queries}.

\subsection{Inferential and differential privacy}

\subsubsection{Inferential privacy}
\label{sec:inferential-privacy}
Inferential privacy \cite{dalenius,inferential-privacy} is the notion that no information about an individual should be learnable with access to a database that could not be learnt without any such access. In a series of important results \cite{revealing-while-privacy,dp-calibrating-noise,differential-privacy}, it was established that such an absolute privacy goal is impossible to achieve if the adversary has access to arbitrary auxiliary information. More importantly, it was observed that individuals' inferential privacy is violated even when they do not participate in the database, because information about them could be leaked by correlated information of other participating individuals.

\subsubsection{Differential privacy}
\label{sec:diffprivacy}
In the wake of the above results, the notion of \emph{differential privacy} was developed \cite{differential-privacy} to allow analysts extract meaningful distributional information from statistical databases while minimising the \emph{additional} privacy risk that each individual incurs by participating in the database. Note that differential privacy is a considerably weaker notion than inferential privacy as reconstruction attacks described in Section \ref{sec:query-restrict} or other correlation attacks can infer a lot of non-identifying information from differentially private databases too.

Mechanisms for differential privacy add noise to the answers depending
on the \emph{sensitivity} of the query. 
In this sense, there is an inherent utility
versus privacy tradeoff.
Differentially private mechanisms possess composability
properties. Thus, privacy degrades gracefully when multiple queries
are made to differentially private databases. However, this alone may
not protect against an attacker making an arbitrary number of
queries. For example, the reconstruction attacks mentioned in Section
\ref{sec:query-restrict} prevent many differentially private
algorithms from answering a linear (in the
number of rows) number of queries \cite{dp-exponential-queries-theory}. For specific
types of queries though, e.g., predicate queries, sophisticated
noise-addition techniques \cite{dp-exponential-mechanism} can be used
to maintain differential privacy while allowing for an exponential
number of queries \cite{dp-exponential-queries-theory,
  dp-exponential-queries-efficient}. 

\subsubsection{Group and societal privacy}
Differentially private mechanisms also degrade gracefully with respect
to group privacy as the group size increases. These guarantees 
may not be enough for policymakers who must protect the profile of
specific communities constituting a sizable proportion of the
population. The ability of an
adversary to manipulate and influence a community even without
explicitly identifying its members is deeply problematic, as
demonstrated by episodes like Cambridge Analytica
\cite{cambridge-analytica}. Therefore, the goal of modern private data
analysis should not be limited to protecting only individual privacy,
but also extend to protecting sensitive
aggregate information.

\subsubsection{A note on non-statistical databases}

Due to the inherently noisy nature of differentially private mechanisms, they are not suitable for any
non-statistical uses, e.g., financial transactions, electronic health
records, and password management. Privacy mechanisms for
such use-cases must prevent misuse of data for malicious purposes such
as illegal surveillance or manipulation, without hampering the
legitimate workflows. 

The difficulties with differential privacy, and the impossibility of protection against inferential privacy violations, suggest that
privacy protection demands that there should be no illegal access or processing in the first place.

\subsection{Purpose limitation}

\subsubsection{Program analysis techniques}

These check whether a given code-base uses
personal data in accordance with a given privacy policy
\cite{legalease,data-capsule,language-based-enforcement}. Privacy
policies are expressed in known 
formal languages
\cite{epal,p3p}. A compiler verifies, using standard information flow analysis \cite{information-flow-analysis} and model-checking
techniques \cite{model-checking}, if a candidate program
satisfies the intended privacy policy. In order to enforce various
information flow constraints these techniques rely on manual and often tedious
tagging
of variables, functions and users with security classes and verify if
information 
does not flow from items
with high security classes 
to items with low security classes.
%

\subsubsection{Purpose-based access control} These techniques define purpose hierarchies and specify purpose-based access-control mechanisms \cite{hippocratic-db,purbac,byunli}. However, they typically identify purpose with the role of the data requester and therefore offer weak protection from individuals claiming wrong purposes for their queries.

\subsubsection{Formalisation of purpose}

Jafari et al. \cite{purpose-semantic-foundation} formalise purpose as
a relationship between actions in an action graph. Hayati et
al. \cite{language-based-enforcement} express purpose as a security
class (termed by them as a ``principal'') and verify that data
collected for a given purpose does not flow to functions tagged with a
different purpose. Tschantz et al. \cite{formalising-purpose} state
that purpose violation happens if an action is \emph{redundant} in a
plan that maximises the expected satisfaction of the allowed purpose.
%
However, enforcement of these models still relies on fine-grained
tagging of code blocks, making them tedious, and either a compiler-based verification or post-facto
auditing, making them susceptible to insider attacks that bypass the checks.

\subsection{Secure remote execution}
\label{ssec:sre}

Secure remote execution refers to the set of techniques wherein a
client can outsource a computation to a remote party such that the
remote party does not learn anything about the client's inputs or intermediate results. 

\subsubsection{Homomorphic encryption}
Homomorphic encryption (HE) schemes compute in the ciphertext space of
encrypted data by relying on the additive or multiplicative
homomorphism of the underlying encryption scheme 
\cite{rsa,elgamal,paillier}. 
Designing an encryption
scheme that is both - which is required for universality - 
is
challenging. Gentry \cite{fhe} gave the first theoretical fully
homomorphic encryption (FHE) scheme. 
Even though
state-of-the-art FHE schemes and implementations have considerably
improved upon Gentry's original scheme, the performance of these
schemes is still 
far from any practical deployment
\cite{survey-fhe}. Functional encryption (FE) \cite{functional-encryption} schemes have similar
objectives, with the crucial difference that FE schemes let the remote
party learn the output of the computation, whereas FHE schemes compute
encrypted output, which is decrypted by the client.

\subsubsection{Secure multiparty computation}
Secure multiparty computation (SMC) - originally pioneered by Yao
through his garbled circuits technique \cite{yao-garbling} - allows
multiple parties to compute a function of their private inputs such
that no party learns about others' private inputs, other than what the
function's output reveals. SMC requires clients to express the
function to be computed as an encrypted circuit and send it to the
server alongwith encrypted inputs; the server needs to evaluate the
circuit by performing repeated decryptions of the encrypted gates. As
a result, SMC poses many challenges in its widespread adoption -
ranging from the inefficiencies introduced by the circuit model itself
to the decryption overhead for each gate evaluation, even as
optimisations over the last two decades have considerably improved the
performance and usability of SMC \cite{smc-hw-survey}.

However, HE, FE and SMC based schemes  involve significant application
re-engineering and may offer reduced functionality in practice.

\subsection{Hardware-based security}
\label{ssec:hardware}

In recent times, secure remote execution is increasingly being
realised not through advances in cryptography but through advances in
hardware-based security. This approach commoditises privacy-preserving
computation, albeit at the expense of a weakened trust model, i.e.,
the increased trust on the hardware manufacturer.

\subsubsection{Intel SGX} 
\label{ssec:hardware-sgx}
Intel Software Guard Extensions (SGX) \cite{sgx} 
implements access control in the CPU 
to provide
confidentiality and integrity to the executing program. At the heart
of the SGX architecture lies the notion of an isolated execution
environment, called an \emph{enclave}. An enclave resides in the
memory space of an untrusted application process but access to the
enclave memory and leakage from it are protected by the hardware. 
The following are the main properties of SGX:

\begin{description}

\item[Confidentiality] 
  Information about an
  enclave execution can not leak outside the enclave memory except
  through explicit exit points.

\item[Integrity] 
  Information can not leak into the enclave to tamper with its execution except through explicit entry points.

\item[Remote attestation] For an enclave's execution to be trusted by
  a remote party, it needs to be convinced that $a)$ the contents of
  the enclave memory at initialisation are as per its expectations,
  and $b)$ that confidentiality and integrity guarantees will be
  enforced by the hardware throughout its execution. For this 
  the hardware computes a \emph{measurement}, essentially a hash of
  the contents of the enclave memory and possibly additional user
  data, signs it and sends it over to the remote party \cite{ra}. 
  The remote party verifies the signature and matches the
  enclave measurement with the measurement of a golden enclave it
  considers secure. If these checks pass, the remote party trusts the
  enclave and sends 
  sensitive inputs to it.

\item[Secure provisioning of keys and data] SGX enclaves have secure
  access to hardware random number generators. Therefore, they can
  generate a Diffie-Hellman public/private key pair and keep the
  private key secured within enclave memory. Additionally, the
  generated public key can be included as part of additional user data
  in the hardware measurement sent to a remote verifier during remote
  attestation. These properties allow the remote verifier to establish
  a secure TLS communication channel with the enclave over which any
  decryption keys or sensitive data can be sent. The receiving enclave
  can also seal the secrets once obtained for long-term use such that it can access them even across reboots, but other programs or enclaves cannot.

\end{description}

\subsubsection{Other hardware security mechanisms}
SGX has been preceded by the Trusted Platform Module (TPM)
\cite{tpm}. TPM defines a hardware-based root of trust, which measures
and attests the entire software stack, including the BIOS, the OS and
the applications, resulting in a huge trusted computing base (TCB) as
compared to SGX whose TCB includes only the enclave code. ARM
Trustzone \cite{trustzone} partitions the system into a secure and an
insecure world and controls interactions between the two. In this way,
Trustzone provides a single enclave, whereas SGX
supports multiple enclaves. Trustzone 
has penetrated the mobile world through ARM-based
Android devices, whereas SGX is available for laptops, desktops and
servers.

SGX is known to be susceptible to serious side-channel attacks
\cite{sgx-controlled-channel-attacks, sgx-cache-attacks,
  sgx-memory-side-channel-attacks,sgx-foreshadow}. Sanctum
\cite{sanctum} has been proposed as a simpler alternative that
provides provable protection against memory access-pattern based
software side-channel attacks.
For a detailed review on hardware-based security, we refer the reader to \cite{smc-hw-survey}.

\subsection{Secure databases}
\label{sec:secure-databases}

Stateful secure remote execution requires a secure database and
mechanisms that protect clients' privacy when they perform queries on
them.

\subsubsection{Querying encrypted databases}
The aim of these schemes is to let clients host their data encrypted
in an untrusted server and still be able to execute queries on it
with minimal privacy loss and maximal query expressiveness. One
approach for enabling this is 
searchable encryption schemes,
i.e., encryption schemes that allow searching over ciphertexts
\cite{search-encrypted-data,peks}. Another approach is to add
searchable indexes along with encrypted data, or to use special
property-preserving encryptions to help with searching
\cite{secure-indexes,symmetric-searchable-encryption,hacigumus,cryptdb}. However,
both approaches are susceptible to inference attacks
\cite{access-pattern-attack-se,leakage-abuse-attacks-sse,attacks-dynamic-sse,attacks-cryptdb}
(cf. Sections \ref{sec:anonymisation}, \ref{sec:query-restrict} and \ref{sec:inferential-privacy}). Oblivious RAM
\cite{oblivious-ram-1,oblivious-ram-2} is a useful primitive that
provides read/write access to encrypted memory while hiding all access
patterns, but these schemes require polylogarithmic number of rounds
(in the size of the database) per read/write request.

EnclaveDB \cite{enclavedb} has been recently proposed as a solution
based on Intel SGX. It hosts the entire database within secure enclave
memory, with a secure checkpoint-based logging and recovery mechanism
for durability, thus providing complete confidentiality and integrity
from the untrusted server without any loss in query expressiveness.
 
\subsubsection{Private information retrieval}
Private information retrieval (PIR) is concerned with hiding which
database rows a given user query touches - thus protecting user intent
- rather than encrypting the database itself. Kushilevitz and
Ostrovsky \cite{computational-pir} demonstrated a PIR scheme with
communication complexity $O(n^{\epsilon})$, for any $\epsilon > 0$,
using the hardness of the quadratic residuosity problem. Since then,
the field has grown considerably and modern PIR schemes boast of
$O(1)$ communication complexity \cite{constant-pir}. Symmetric PIR
(also known as oblivious transfer), i.e., the set of schemes where
additionally users cannot learn anything beyond the row they
requested, is also an active area of research.

\section{Operationalisation using trusted executables and regulatory architecture}
\label{sec:trusted-executables}

As is evident from the discussion in the previous section, none of the techniques by themselves are adequate for privacy protection. In particular,  none are effective against determined insider attacks without regulatory oversight. Hence we need an overarching architectural framework based on regulatory control over data minimisation, authorisation, access control and purpose limitation.  In addition, since the privacy and fairness impacts of modern AI techniques \cite{fairml-book} are impossible to determine automatically, the regulatory scrutiny of data  processing programs must have a best effort manual component. Once approved, the architecture must prevent any alteration or purpose extension without regulatory checks.

In what follows we present an operational architecture for privacy-by-design.
We assume that all databases and the associated computing environments are under physical control of the data controllers, and the online regulator has no direct physical access to it. We also assume that the data controllers and the regulators do not collude.

We illustrate our conceptual design through an example of a privacy-preserving electronic health record (EHR) system. EHRs can improve  quality of healthcare significantly by providing improved access to patient records to doctors, epidemiologists and policymakers. However, the privacy concerns with them are many, ranging from the social and psychological harms caused by unwanted exposure of individuals' sensitive medical information, to direct and indirect economic harms caused by the linkage of their medical data with data presented to their employers, insurance companies or social security agencies. Building effective EHRs while minimising privacy risks is a long standing design challenge.


\subsection{Trusted executables}

\begin{figure*}[t]
    \centering
    \begin{tabular}{c c}
    \begin{tabular}{l}
	    \includegraphics[width=12cm]{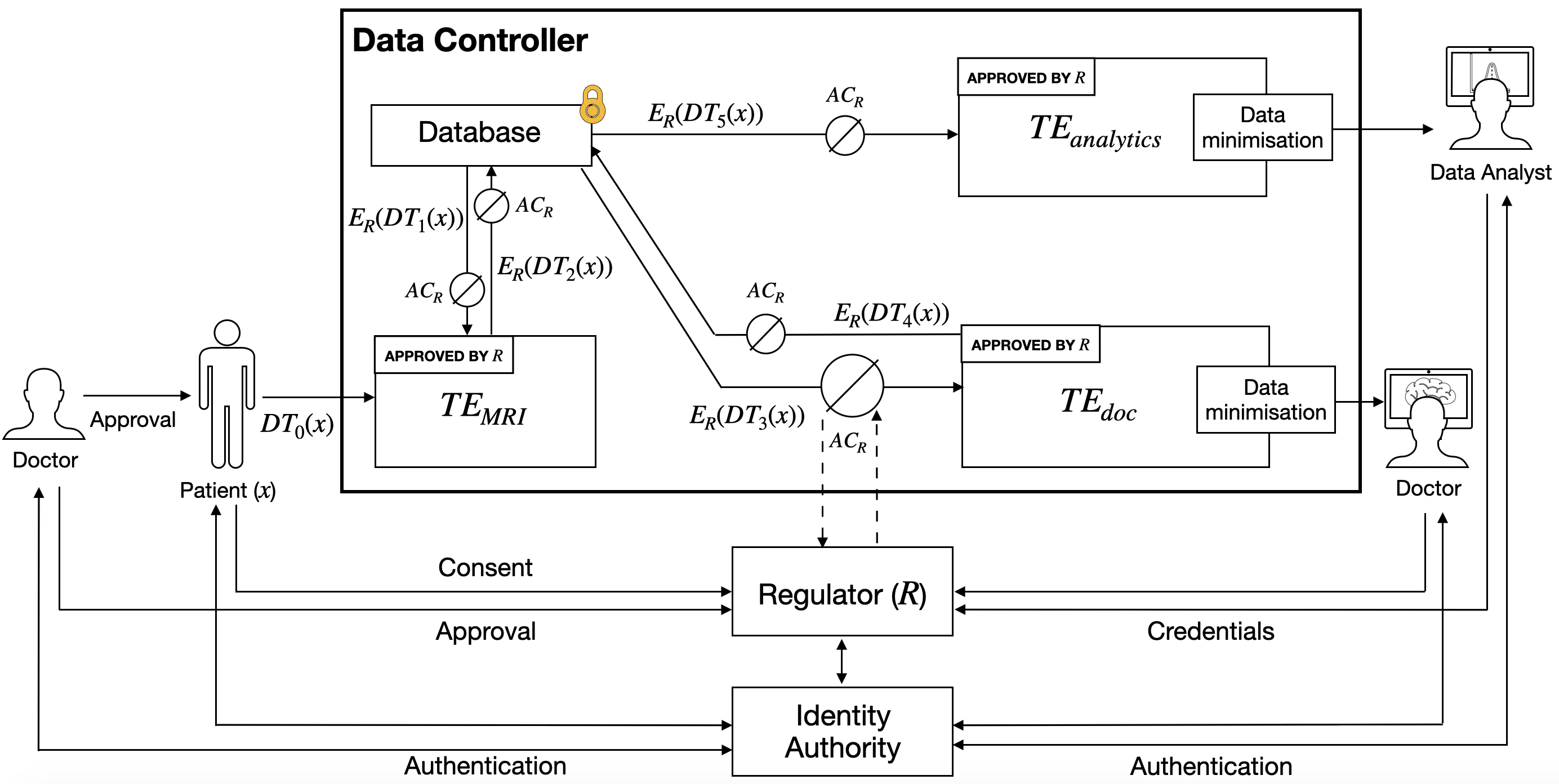}
    \end{tabular} &
    \begin{tabular}{l}
	    \includegraphics[width=4cm]{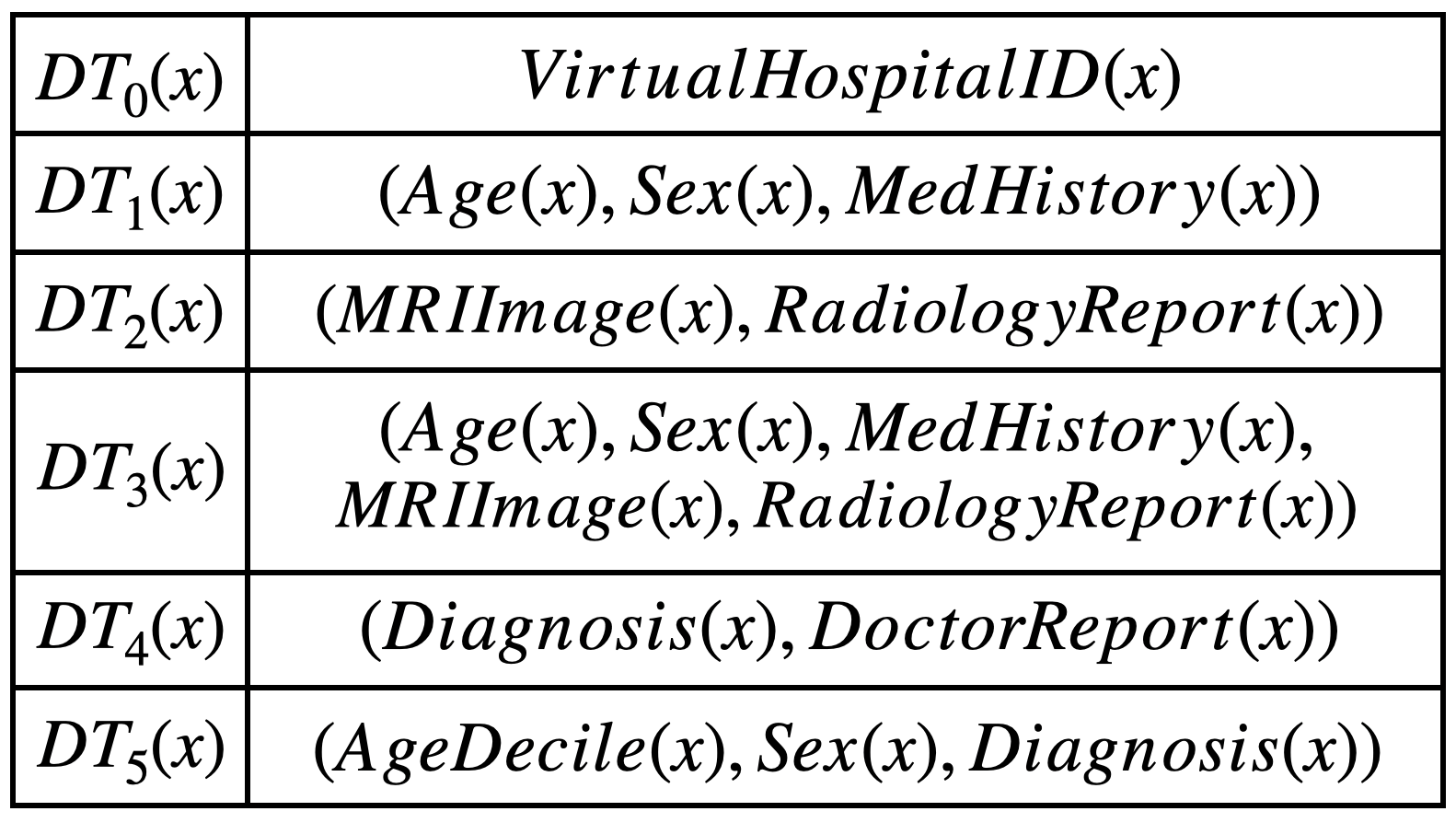}
    \end{tabular}
    \end{tabular}
    \caption{\small An illustration of the architecture of trusted executables using an example involving an EHR database, a patient, an MRI imaging station, a doctor and a data analysis station. TEs marked ``{\textbf{\footnotesize APPROVED BY }$R$}''  are pre-audited and pre-approved by the regulator $R$. $E_R(\cdot)$ represents a regulator-controlled encryption function and $AC_R$ represents online access control by regulator $R$. $DT_i(x)$ represent various data types parametrised by the patient $x$ (as explained in the right-hand side table). In particular, $VirtualHospitalID(x)$ represents the hospital-specific virtual identity of the patient. The regulator checks the consents, approvals and other static rules regarding data transfer at each stage of online access control.}
    \label{fig:trusted-executables}
\end{figure*}

We propose \emph{trusted executables} (TE) as the fundamental building blocks for privacy-by-design. We introduce them in the abstract, and discuss some possibilities for their actual realisation in Section \ref{sec:feasibility}. TEs are data-processing programs, with explicit input and output channels, that are designed by the data controllers but are examined, audited, and approved   by appropriate regulators. TEs execute in controlled environments on predetermined data types with prior privacy risk assessment, under online regulatory access control. The environment ensures that only approved TEs can operate on data items. In particular, all data accesses from the databases, and all data/digest outputs for human consumption, can only happen through the TEs. We prescribe the following main properties of the TEs:
\begin{enumerate}
     \item \emph{Runtime environment}:  TEs are approved by regulators. They execute in the physical infrastructure of the data controllers but cannot be modified by them. 
    \item \emph{Authentication}: A regulator can authenticate related TEs during runtime, and verify that indeed the approved versions are running.
    \item \emph{Integrity}: There is no way for a malicious human or machine agent, or even for the data controller, to tamper with the execution of a TE other than by sending data through the TE's explicit input channels.
    \item \emph{Confidentiality}: There is no way for any entity to learn anything about the execution of a TE other than by reading data written at the TE's explicit output channels. 
    All data accesses and output can only happen through TEs.
\end{enumerate}
Besides, all TEs should be publicly available for  scrutiny. The above properties  allow a regulator to ensure that a TE is untampered and will conform to the limited purpose identified at the approval stage.  

As depicted in Figure \ref{fig:trusted-executables}, a data agent - for example, a hospital  - interacts with databases or users only through pre-approved TEs, and  direct accesses are prevented. All data stores and communication messages are encrypted using a regulator-controlled encryption scheme  to prevent any information leakage in transit or storage. The data can be decrypted only inside the TEs under regulated access control. The regulator provisions decryption keys securely to the TE to enable decryption after access is granted. The regulator allows or denies access, online,  based on the authentication of the TE and the incoming data type, consent and approval checks as required, and the credential authentication of any human consumers of output data (e.g., the doctor(s) and  data analysts).   All sink TEs  - i.e., those that output data directly for consumption by a human agent - are pre-audited to employ appropriate data minimisation  before sending data to their output channels.  Note that extending the TE architecture to the doctors' terminals and the imaging stations ensures that the data never crosses the regulated boundary and thus enables purpose limitation.

In the above example  an independent identity authority  issues credentials and engages in a three-way communication to authenticate individuals who present their virtual identities to the regulator. An individual can use a master {\em Health id} to generate hospital-specific or doctor-specific unlinkable anonymous credentials. Only a {\em health authority} may be allowed to link identities across hospitals and doctors in a purpose-limited way under regulated access control.

\subsection{Regulatory architecture}



\begin{figure*}[t]
    \centering
    \includegraphics[width=14cm]{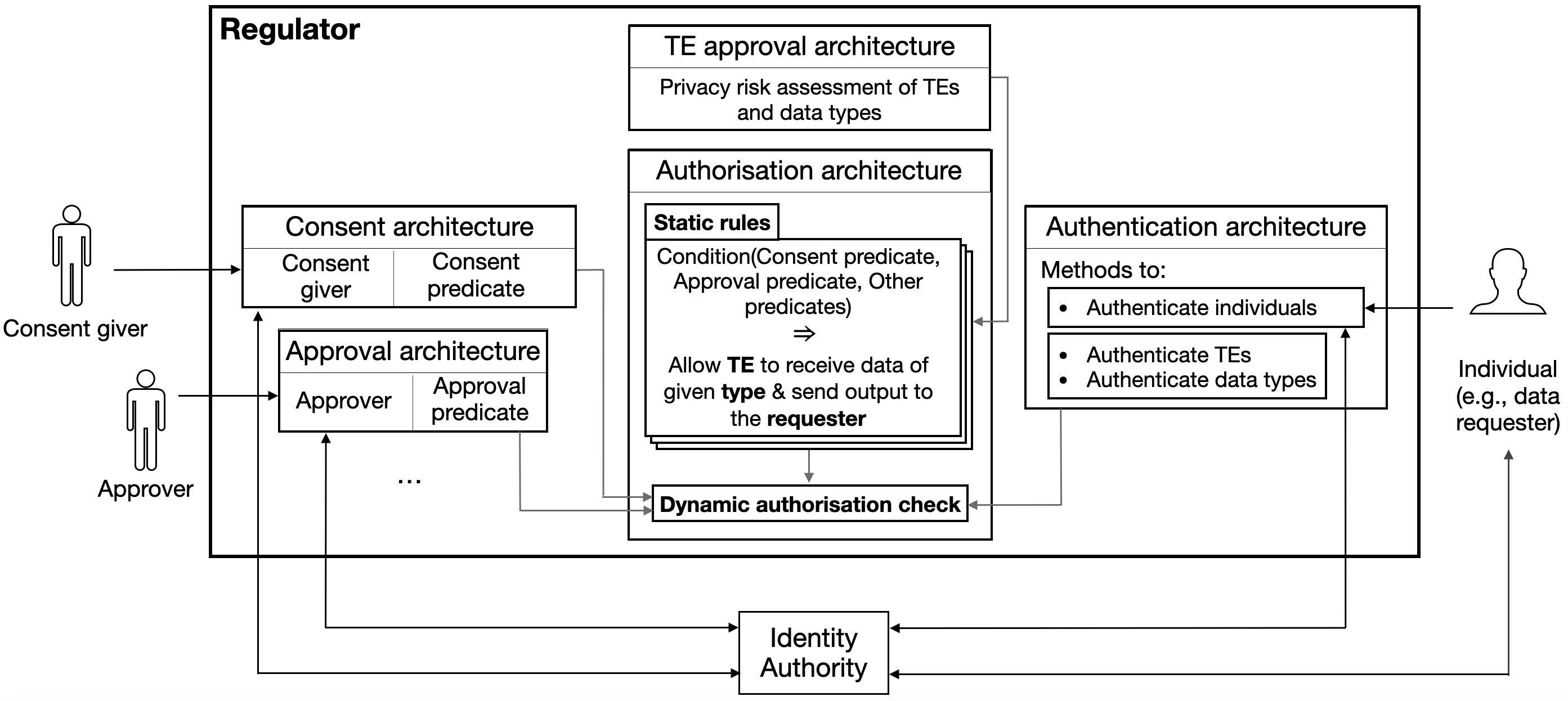}
    \caption{\small The regulatory architecture.}
    \label{fig:regulator-architecture}
\end{figure*}

\begin{figure*}[t]
    \centering
    \includegraphics[width=13cm]{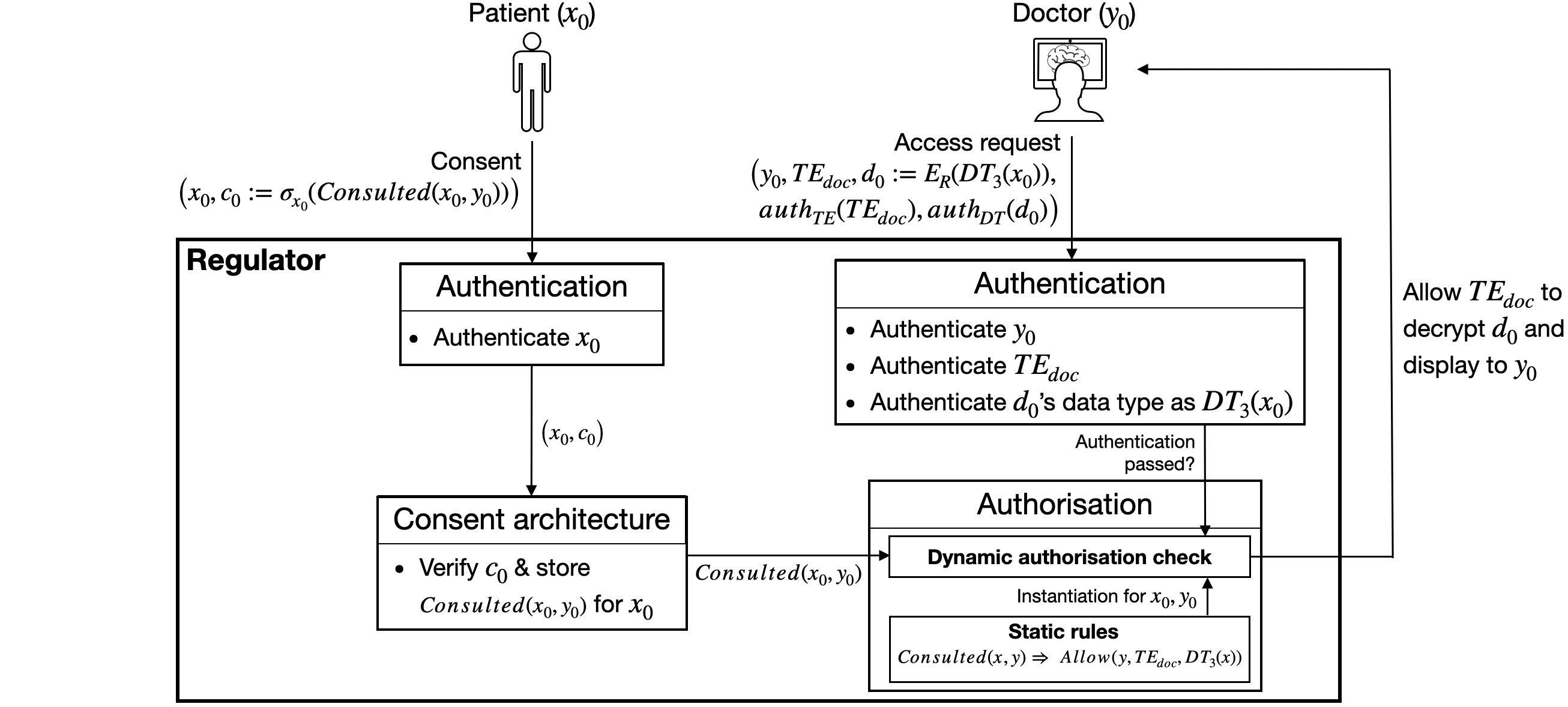}
    \caption{\small An example control flow diagram depicting the regulatory architecture. $x_0$ and $y_0$ represent virtual identities of the patient and the doctor, respectively. $\sigma_{x_0}(\cdot)$ represents digital signature by patient $x_0$. $auth_{TE}(\cdot)$ represents authentication information of the TE and $auth_{DT}(\cdot)$ represents authentication information of the supplied data's type. Individuals are authenticated by verifying their virtual identities.}
    \label{fig:ehr-flow}
\end{figure*}

We depict the regulatory architecture in Figure \ref{fig:regulator-architecture}. The first obligation of the regulator is to audit and approve the TEs designed by the data controllers. During this process, the regulator must assess the legality of the data access and processing requirements of each TE, along with the privacy risk assessment of its input and output data types. In case a TE is an AI based data analytics program, it is also an obligation of the regulator to assess its fairness and the potential risks of  discrimination \cite{fairml-book}. Before approving a TE, the regulator also needs to verify that the TE invokes a callback to the regulator's online interface before accessing a data item and supplies appropriate authentication information, and that it employs appropriate encryption and data minimisation mechanisms at its output channels. Finally, the regulator needs to put in place a mechanism to be able to authenticate the TE in the data controller's runtime environment.

 The second obligation of the regulator is to play an online role in authorising data accesses by the TEs. The authorisation architecture has both a static  and a dynamic component. The static authorisation rules typically capture the relatively stable regulatory requirements, and the dynamic component typically captures the fast-changing online context, mainly due to consents and approvals. Specifically, each static authorisation rule takes the form of a set of pre-conditions necessary to grant access to a TE the data of a given type; and, in case of sink TEs, to output it to a requester. The design of these rules is governed by regulatory requirements and the privacy risk assessment of TEs and data types. The rules are typically parametric in nature,  allowing specification of constraints that provide access to a requester only if the requester can demonstrate some specific relationship with the data individual (e.g., to express that only a doctor consulted by a patient can access her data).

The pre-conditions of the authorisation rules may be based on consent of data individuals, approvals by authorities or even other dynamic constraints (e.g., time-bound permissions). The consent architecture must be responsible for verifying signatures on standardised consent APIs from consent givers and recording them as logical consent predicates. The regulator, when designing its authorisation rules, may use a simple consent - for example, that a patient has wilfully consulted a  doctor - to invoke a set of rules  to protect the individual's  privacy under a legal framework, rather than requiring individuals to self-manage their privacy. 

Similar to the consent architecture, the approval architecture for data access must record standardised approvals from authorities as logical approval predicates. An approval from an authority may also be provided to an individual instead of directly to the regulator,  as a blind signature against a virtual identity of the individual known to the approver, which should be transformed by the individual to a signature against the virtual identity known to the data controller and the regulator. This, for example, may enable a patient to present a self generated virtual identity to a doctor or a hospital instead of her universal {\em Health id}.

The regulator also requires an authentication architecture. First, it needs to authenticate individuals, i.e., consent givers, approvers and data requesters, by engaging in a three-way communication with an identity authority which may be external to both the data controller and the regulator. Second, it needs to authenticate TEs in order to be able identify the access requests as originating from one of the pre-approved TEs. Third, it needs to authenticate data types, i.e., identify the underlying type of the TE's encrypted input data.

The consent/approval predicates and the authentication information flow to the dynamic authorisation module, which can instantiate the static authorisation rules with the obtained contextual information to determine, in an online fashion, if access should be allowed to the requesting TE. If yes, then it must also provision decryption keys to the TE securely such that only the TE can decrypt. The keys can be securely provisioned to the TE because of the authentication, integrity and confidentiality properties, and by the fact that approved TEs must never output the obtained decryption keys. 

An example control-flow diagram depicting the regulatory access control in a scenario where a doctor is trying to access the data of a patient who consulted them is shown in Figure \ref{fig:ehr-flow}.

\section{Discussion on feasibility}
\label{sec:feasibility}

Several existing techniques can be useful for the proposed architecture, though some techniques may need strengthening.

{\bf Trusted executables}  can be implemented most
  directly on top of trusted hardware primitives such as Intel SGX or
  ARM Trustzone where authentication of TEs is carried out by remote
  attestation. 
{\bf Secure provisioning of keys and data} to TEs can be done in case of Intel
  SGX as per Section \ref{ssec:hardware-sgx}. However, since SGX includes only the CPU in its TCB, it presents challenges in porting AI applications that run on GPUs for efficiency. Graviton \cite{graviton} has been recently proposed as an efficient hardware architecture for trusted execution environments on GPUs.

In our architecture, TEs  fetch or update information
from {\bf encrypted databases}.
  This may be implemented using special indexing data structures, or may  involve search over encrypted data, where the TEs
  act as clients and the database storage acts as the
  server. Accordingly, techniques from Section
  \ref{sec:secure-databases} can be used. 
  Since
  the TEs never output data to agents unless deemed legitimate by the
  regulator, the inferential attacks identified with these schemes in
  Section \ref{sec:secure-databases} have minimal impact. For added
  security, EnclaveDB \cite{enclavedb}, which keeps the entire
  database in secure enclave memory, can be used. EnclaveDB has been evaluated on standard database benchmarks TPC-C \cite{tpcc} and TATP \cite{tatp} with promising results.

For {\bf authentication of data types} 
  messages may be encrypted using an ID-based encryption
  scheme, where the concrete runtime type of the message acts as the
  textual ID and the regulator acts as the trusted third party (see
  Section \ref{ssec:encryption}). The receiver TE can send the
  expected plaintext type to the regulator as part of its access
  request. The regulator should provision the decryption key for the
  ID representing the requested type only if the receiver TE is
  authorised to receive it as per the dynamic authorisation
  check. Note that authentication of the received data type is
  implicit here, as a TE sending a different data type in its access
  request can still not decrypt the incoming data.

{\bf Data minimisation} for consents and approvals
  based on {\bf virtual identities} is well-established from Chaum's original
  works \cite{vid,anoncred}. Individuals should use their
  purpose-specific virtual identities with organisations, as opposed
  to a unique master identity. 
  To prevent cross-linking of identities, {\bf anonymous credentials}
  may be used.
In some cases, individuals' different {\bf virtual identities may need to
be linked} by a central authority to facilitate data analytics or
inter-organisation transactions. This should be done under strict
regulatory access control and purpose limitation.

 Modern type systems can conveniently
  express the complex parametric constraints in the rules in the {\bf authorisation
  architecture}. Efficient type-checkers and logic engines exist that could
  perform the dynamic authorisation checks.

{\bf Approval of TEs}  needs to be largely
  manual as the regulator needs to evaluate the legitimacy and privacy risks associated with the proposed data
  collection and processing activity. However, techniques from program analysis
  may help with specific algorithmic tasks, such as checking
  if the submitted programs adhere to the structural requirement of
  encrypting data items with the right type at their outgoing
  channels.

We require the {\bf regulatory boundary to be extended} even to agent machines,
which must also run TEs so that data they obtain is not
  repurposed for something else. However, when a TE at an authorised
  agent's machine outputs data, it could be intercepted by malicious
  programs on the agent's machine 
  leading to  purpose violation.
  Solutions from the DRM literature may  be used to prevent this. In particular,
  approaches that directly encrypt data for display devices may be
  useful \cite{hdcp}. We note that this still does not protect the
  receiving agent from using more sophisticated mechanisms to copy
  data (e.g., by recording the display using an external device). However,
  violations of this kind are largely manual in nature and ill-suited
  for large-scale automated attacks. 

Finally, we need {\bf internal processes at the regulatory authority} itself
 to ensure that its actual operational code
  protects the various decryption keys and provides access to TEs as
  per the approved policies. To this end, the regulator code may
  itself be put under a TE and authenticated by the regulatory
  authority using remote attestation. Once authenticated, a master
  secret key may be provisioned to it using which the rest of the
  cryptosystem may bootstrap.

\section{Additional case studies}
\label{sec:case-studies}

In this section, we present two additional case studies to showcase the applicability of our architecture in diverse real-world scenarios. 

\subsection{Direct Benefit Transfer}
\label{ssec:dbt}

Direct Benefit Transfer (DBT) \cite{dbt} is a Government of India scheme  to transfer subsidies to  citizens' bank accounts under various welfare schemes. Its primary objective is to bring transparency and reduce leakages in public fund disbursal. The scheme design is based on India's online national digital identity system Aadhaar \cite{aadhaar}. All DBT recipients  have their Aadhaar IDs  linked to their bank accounts to receive benefits.

\begin{figure*}[t]
    \centering
    \includegraphics[width=10cm]{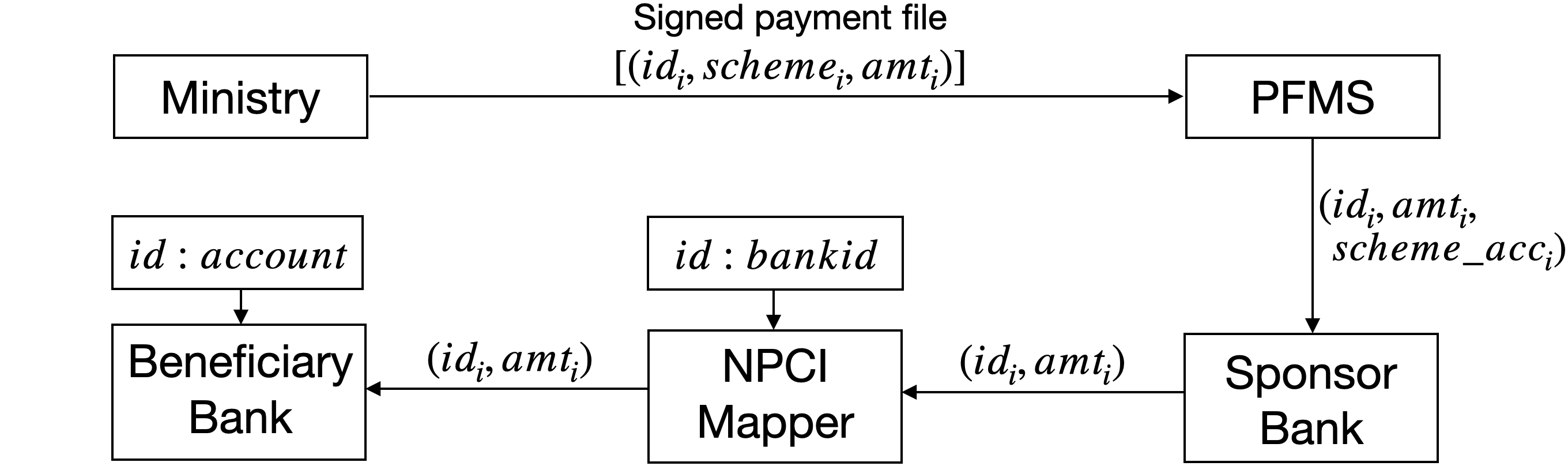}
    \caption{\small A simplified schematic of direct benefit transfer as it exists today.}
    \label{fig:dbt-current}
\end{figure*}

\begin{figure*}[t]
    \centering
    \includegraphics[width=12cm]{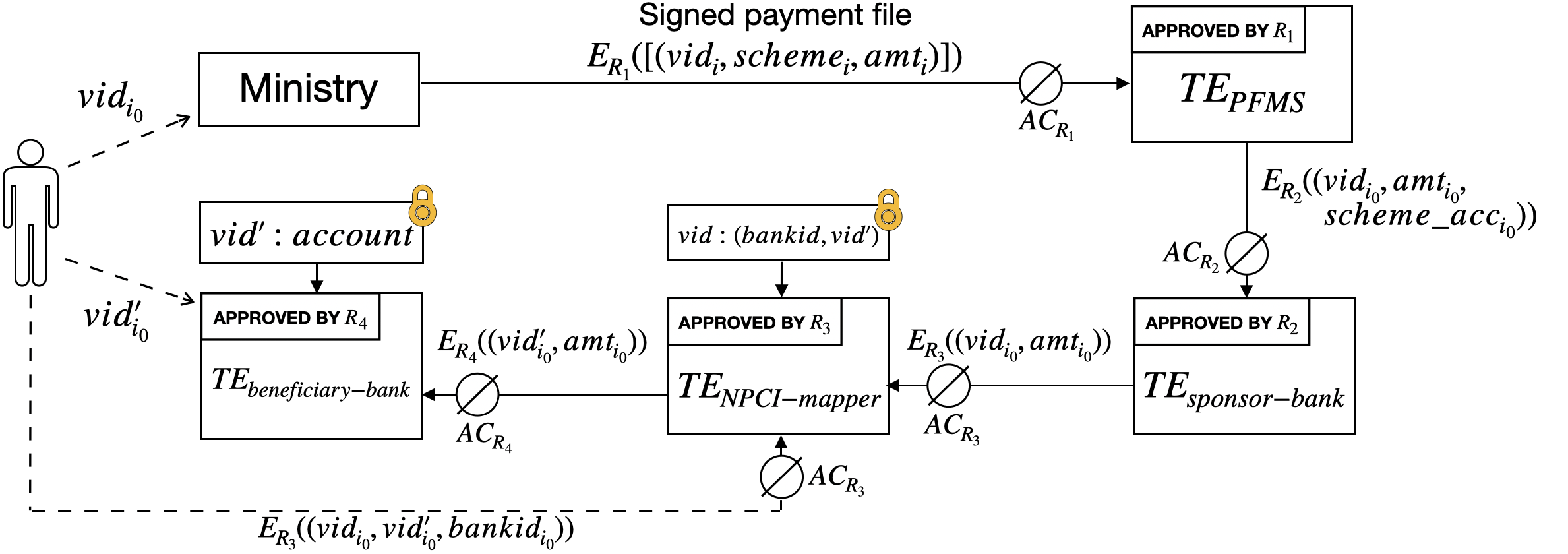}
    \caption{\small Our proposal for privacy-preserving direct benefit transfer. $vid_{i_0}$ and $vid'_{i_0}$  represent $i_0$'s DBT-specific and Bank-specific virtual identities, respectively.  Dashed arrows represent one-time onboarding steps. $R_1$, $R_3$, $R_2$ and $R_4$ represent a DBT regulator,  a centralised financial regulator,  and internal regulators of the sponsoring  and the  beneficiary banks, respectively.}
    \label{fig:dbt-proposal}
\end{figure*}

Figure \ref{fig:dbt-current} shows a simplified schematic of the scheme that exists today \cite{dbt1}. A ministry official initiates payment  by generating a payment file detailing the Aadhaar IDs of the DBT recipients, the welfare schemes under which payments are being made and the amounts to be transferred. The payment file is then signed and sent to a centralised platform called the Public Financial Management System (PFMS). PFMS hosts the details of various DBT schemes and is thus able to initiate an inter-bank fund transfer from the bank account of the sponsoring scheme to the bank account of the beneficiary, via the centralised payments facilitator NPCI (National Payments Corporation of India). NPCI maintains a mapping of citizen's Aadhaar IDs to the banks hosting their DBT accounts. This mapping allows NPCI to route the payment request for a given Aadhaar ID to the right beneficiary bank. The beneficiary bank internally maintains a mapping of its customers' Aadhaar IDs to their bank account details, and is thus able to transfer money to the right account.

As DBT payments are primarily directed towards people who need benefits, precisely because they are structurally disadvantaged, their DBT status must be protected from future employers, law enforcement, financial providers etc., to mitigate discrimination and other socio-economic harms coming their way. Further, since DBT relies on the recipients' national Aadhaar IDs, which are typically linked with various other databases, any leakage of this information makes them directly vulnerable. Indeed, there are reports that bank and address details of millions of DBT recipients were leaked online \cite{cis}; in some cases this information was misused to even redirect DBT payments to unauthorised bank accounts \cite{airtel}.

We illustrate our approach for a privacy-preserving DBT in Figure \ref{fig:dbt-proposal}. In our proposal, DBT recipients use a virtual identity for DBT that is completely unlinkable to the virtual identity they use for their bank account. They may generate these virtual identities - using suitably designed simple and intuitive interfaces - by an anonymous credential scheme where the credentials are issued by a centralised identity authority. Additionally, they provide the mapping of the two virtual identities, along  with the bank name, to the NCPI mapper. This information is provided encrypted under the control of the financial regulator $R_3$ such that only the NPCI mapper TE can access it under $R_3$'s online access control. This mechanism allows the NPCI mapper to convert payment requests against DBT-specific identities to the target bank-specific identities, while maintaining the mapping private from all agents. Regulator-controlled encryption of data in transit and storage and the properties of TEs allow for an overall privacy-preserving DBT pipeline.

Note that data flow is controlled by different regulators along the DBT pipeline, providing a distributed approach to privacy protection. PFMS is controlled by a DBT regulator; NPCI mapper is controlled by a financial regulator, and the sponsor and beneficiary banks are controlled by their respective internal regulators. 

\subsection{Contact tracing}
\label{ssec:contact-tracing}

\begin{figure*}[ht]
    \centering

    \begin{subfigure}{.9\linewidth}
      \centering
      \includegraphics[width=.6\linewidth]{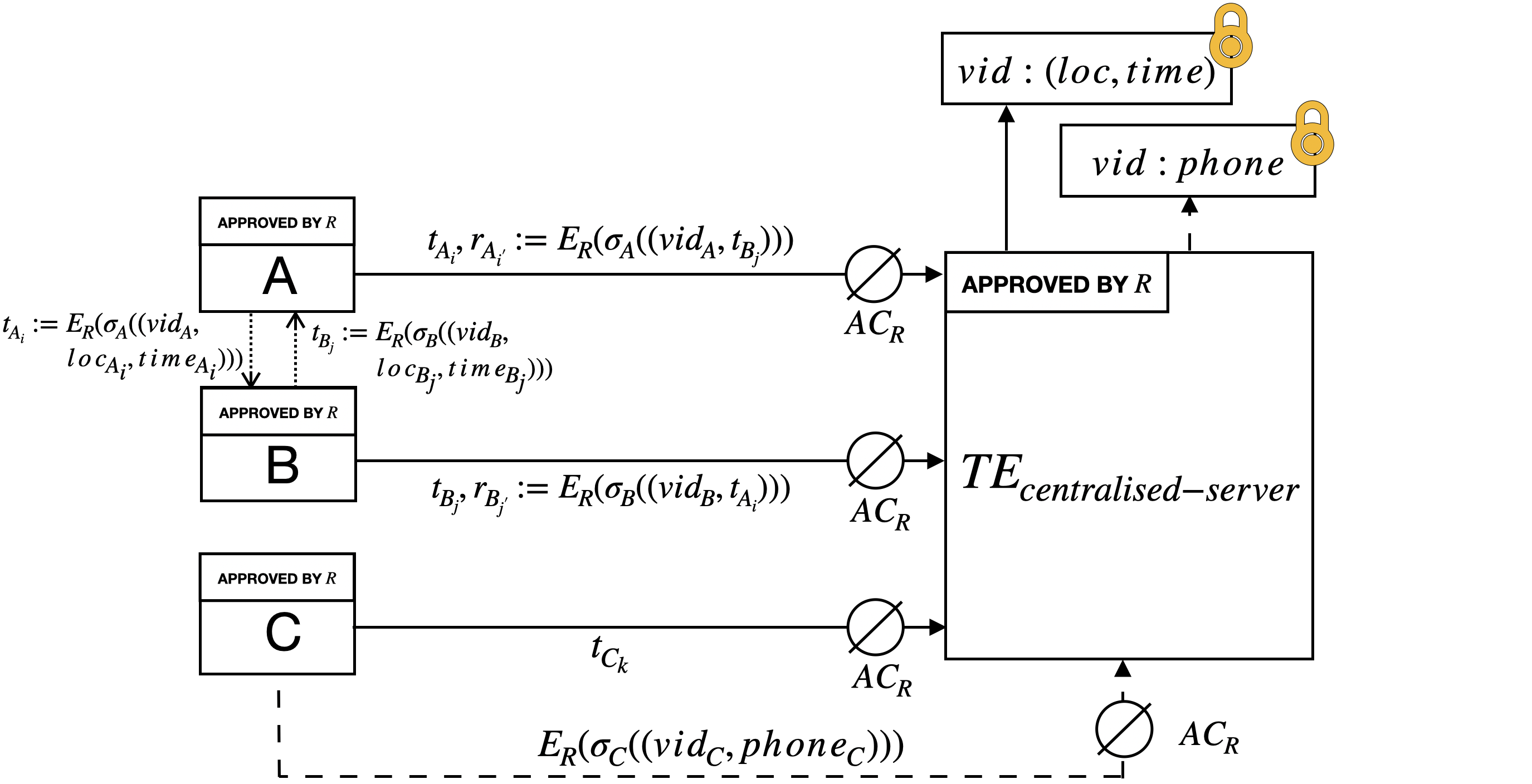}
      \caption{\small Collecting spatiotemporal information. $A$ and $B$ come in contact via BLE, as denoted by the dotted arrows. $C$ does not come in contact with $A$ or $B$ via BLE but is spatially close within a time window, as per GPS data. $vid_{X}$ represents the virtual identity of agent $X$; $loc_{X_i}$ represents $X$'s $i$-th recorded location; $time_{X_i}$ represents its $i$-th recorded time. $t_{X_i}$ represents $i$-th token generated by $X$; $r_{X_i}$ represents $i$-th receipt obtained by $X$; $\sigma_X()$ represents signing by $X$. Dashed arrows represent one-time registration steps (illustrated only for $C$).}
      \label{fig:contact-tracing-1}
    \end{subfigure}

    \bigskip

    \begin{subfigure}{.9\linewidth}
      \centering
      \includegraphics[width=.95\linewidth]{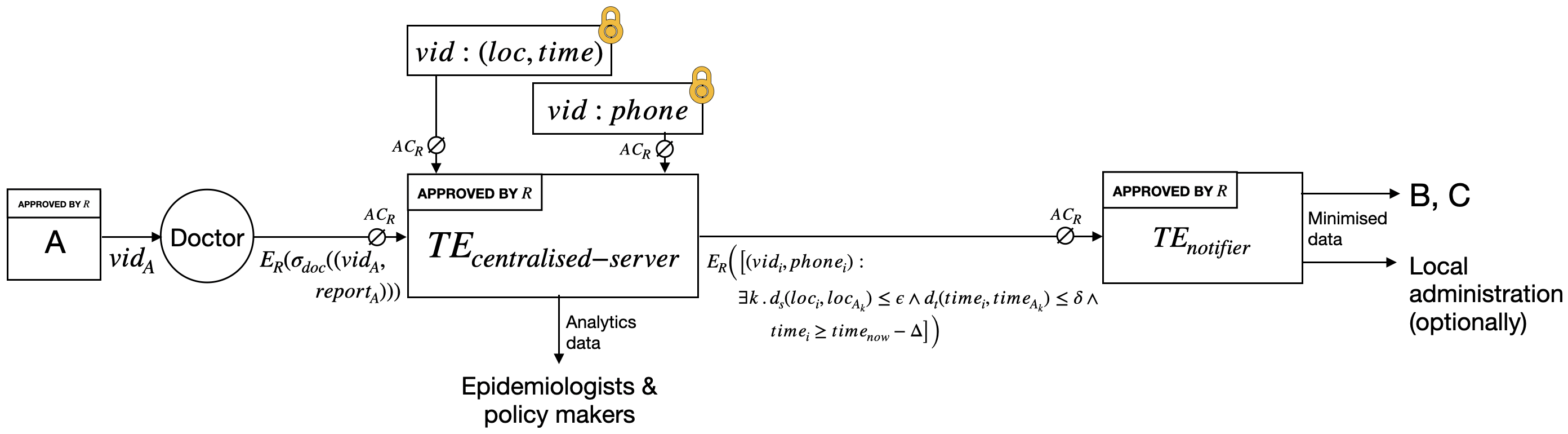}
      \caption{\small Tracing the contacts of infected individuals. $A$ gets infected, as certified by the doctor's signature $\sigma_{doc}$ on $A$'s virtual identity $vid_A$ and medical report $report_A$. $d_s$ and $d_t$ respectively represent chosen spatial and temporal distance functions and $\epsilon$ and $\delta$ the corresponding thresholds, as per the disease characteristics. $\Delta$ represents the infection window, the time during which $A$ might have remained infectious. $time_{now}$ represents the time when the query was executed.}
      \label{fig:contact-tracing-2}
    \end{subfigure}
    
    \caption{\small Contact tracing}
    \label{fig:contact-tracing}
\end{figure*}

There have been a plethora of attempts recently from all over the world towards electronic app-based contact tracing for COVID-19  using a combination of GPS and Bluetooth \cite{krishnan-covid-china,covid-south-korea,tracetogether,aarogyasetu,raskarapp,canettiapp,epione,applegooglecovidapp,mitpact}. Even keeping aside the issue of their effectiveness,  some serious privacy concerns have been raised about such apps. 

In most of these apps the smartphones exchange anonymous tokens when they are in proximity, and each phone keeps a record of the sent and received  tokens. When an individual is infected - signalled either through a self declaration or a testing process - the tokens are uploaded to a central service.

There are broadly two approaches to contact tracing:
\begin{enumerate}
\item those involving a trusted central authority that can decrypt the tokens and, in turn, alert individuals and other authorities about potential infection risks \cite{krishnan-covid-china,covid-south-korea,tracetogether,aarogyasetu}. Some of these apps take special care to not upload any information about individuals who are not infected.
\item those that assume that the central authority is untrusted and use  privacy preserving computations  on user phones to alert individuals about their potential risks of infection  \cite{raskarapp,canettiapp,epione,applegooglecovidapp,mitpact}. The central service just facilitates access to anonymised sent tokens of infected individuals and cannot itself determine the infection status of anybody.
\end{enumerate}

The following are the main privacy attacks on contact tracing apps: 1) individuals learning about other individuals as high-risk spreaders, 2) insiders at the central service learning about individuals classified as high risk, 3) exposure of social graphs of individuals, and 4) malicious claims by individuals forcing quarantine on others. See \cite{epione} for a vulnerability analysis of some popular approaches.

The centralised approaches clearly suffer from many of the above privacy risks. While alerting local authorities about infection risks is clearly more effective from a public health perspective, to enable them to  identify  hotspots and make crucial policy decisions, it is mainly the privacy concerns that sometimes motivate the second approach.  Also, it is well known that location data of individuals can be used to orchestrate de-anonymisation attacks \cite{deanon-location}, and hence many of the above approaches adopt the  policy of not using geolocation data for contact tracing despite their obvious usefulness at least in identifying hotspots. In addition, Bluetooth based proximity sensing - which are isolated communication events over narrow temporal windows between two smartphones - is ineffective for risk assessment of indirect transmission through contaminated surfaces, where the virus can survive for long hours or even days. Such risk assessment will require computation of intersection of space-time volumes of trajectories which will be difficult in a decentralised approach. It appears that the privacy considerations have forced many of these approaches to adopt overly defensive decentralised designs at the cost of effectiveness.


In contrast, we propose an architecture where  governments can collect fine-grained location and proximity data of citizens,  but under regulated access control and purpose limitation.  Such a design can support both short-range peer-to-peer communication technologies such as BLE and GPS based location tracking. Additionally, centralised computing can support  space-time intersections.

In Figure \ref{fig:contact-tracing}, we show the design of a \emph{state-mandated} contact-tracing app that, in addition to protecting against the privacy attacks outlined earlier, can also protect against attacks by individuals who may maliciously try to pose as low-risk on the app, for example to get around restrictions (attack 5). 

As before, we require all storage and transit data to be encrypted under a regulator-controlled encryption scheme, and that they be accessible only to pre-approved TEs. We also require the app to be running as a TE on the users' phones (e.g., within a trusted zone on the phone). 

We assume that everyone registers with the app using a phone number and a virtual identity unlinkable to their other identities. Periodically, say after every few minutes, each device records its current GPS location and time. The tuple made up of the registered virtual identity and the recorded location and time is signed by the device and encrypted controlled by the regulator, thus creating an ephemeral ``token'' to be shared with other nearby devices over BLE. When a token is received from another device, a tuple containing the  virtual identity of self and the incoming token is created, signed and stored in a regulator-controlled encrypted form, thus creating a ``receipt''. Periodically, once every few hours, all locally stored tokens and receipts are uploaded to a centralised server TE, which stores them under regulated access control as a mapping between registered virtual identities and all their spatiotemporal coordinates. For all the receipts, the centralised server TE stores the same location and time for the receiving virtual identity as in the token it received, thus modelling the close proximity of BLE contacts.

When a person tests positive, they present their virtual identity to a medical personnel who uploads a signed report certifying the person's infection status to the centralised server TE. This event allows the centralised server TE to fetch all the virtual identities whose recorded spatiotemporal coordinates intersects within a certain threshold, as determined by the disease parameters, with the infected person's coordinates. As the recorded (location, time) tuples of any two individuals who come in contact via BLE necessarily collide in our approach, the virtual identities of all BLE contacts can be identified with high precision. Moreover, virtual identities of individuals who did not come under contact via BLE but were  spatially nearby in a time window as per GPS data are also identified. 

A notifier TE securely obtains the registered phone numbers corresponding to these virtual identities from the centralised server TE and sends suitably minimised notifications to them, and also perhaps to the local administration according to local regulations. The collected location data can also be used independently by epidemiologists and policy makers in aggregate form to help them understand  the infection pathways and  identify areas which need more resources.

Note that attack 1 is protected by the encryption of all sent tokens; attacks 2 and 3 are protected by the properties of TEs and regulatory access control; attack 4 is protected by devices signing their correct spatiotemporal coordinates against their virtual identity before sending tokens or receipts. Attack 5 is mitigated by requiring the app to run within a trusted zone on users' devices, to prevent individuals from not sending tokens and receipts periodically or sending junk data.

\section{Conclusion}

We have presented the design sketch of an operational architecture for privacy-by-design \cite{pbd} based on regulatory oversight, regulated access control, purpose limitation and data minimisation. We have established the need for such an architecture by highlighting limitations in existing approaches and some  public service application designs. We have demonstrated its usefulness with some case studies.


While we have explored the feasibility  of our architecture based on existing techniques in computer science, some of them will definitely require further strengthening.  There also needs to be detailed performance and usability evaluations, especially in the context of large-scale database and AI applications. Techniques to help a regulator assess the privacy risks of TEs also need to be investigated. These are interesting open problems that need to be solved to create practical systems for the future with built-in end-to-end privacy protection.

\pagebreak

\bibliography{pbd}
\bibliographystyle{ieeetr}

\end{document}